\documentclass[12pt]{article}

\usepackage{amsmath,graphicx,amssymb}

\numberwithin{equation}{section}

\begin{document}

\title{Dynamics of polarization buildup by spin filtering}
\author{N.~H.~Buttimore{\ }
         and{\ }
D.~S.~O'Brien\thanks{donie@maths.tcd.ie}
\\[2ex]
School of Mathematics, Trinity College Dublin, Ireland}

\date{\today}

\maketitle

\vspace*{-3ex}

\begin{abstract}

There has been much recent research into polarizing an antiproton beam, instigated by the recent proposal from the PAX (Polarized Antiproton eXperiment) project at GSI Darmstadt.  It plans to polarize an antiproton beam by repeated interaction with a polarized internal target in a storage ring.  The method of polarization by spin filtering requires many of the beam particles to remain within the ring after scattering off the polarized internal target via electromagnetic and hadronic interactions.  We present and solve sets of differential equations which describe the buildup of polarization by spin filtering in many different scenarios of interest to projects planning to produce high intensity polarized beams.  These scenarios are: 1) spin filtering of a fully stored beam, 2) spin filtering while the beam is being accumulated, {\it i.e.}\ unpolarized particles are continuously being fed into the beam, 3) the particle input rate is equal to the rate at which particles are being lost due to scattering beyond ring acceptance angle, the beam intensity remaining constant, 4) increasing the initial polarization of a stored beam by spin filtering, 5) the input of particles into the beam is stopped after a certain amount of time, but spin filtering continues.  The rate of depolarization of a stored polarized beam on passing through an electron cooler is also shown to be negligible.

\end{abstract}

\section{Introduction}
\label{intro}

The availability of a stored polarized antiproton beam offers
 many new opportunities for studying the structure of hadronic matter.
 The collision of polarized protons with polarized antiprotons
 at relatively high luminosity, for example, enables an evaluation
 of the transverse polarization of a quark within a proton.
 Such a determination would provide an important element
 of a QCD description of the partonic structure of a nucleon.
 Transversity can only readily be accessed through a Drell-Yan process
 induced by the scattering of appropriately polarized protons
 and polarized antiprotons \cite{Anselmino:2004ki,Pasquini:2005dk,Pasquini:2006iv,Anselmino:2007fs,Barone:2005pu}.  

Another area of study that would be opened up as a consequence of having an intense beam of polarized antiprotons relates to the time-like electromagnetic form factors of a proton, particularly their phases, since the form factors take complex values in the time-like region above threshold \cite{Gakh:2006rj,Buttimore:2007cv}. The expectations of perturbative QCD could be probed in some detail with such form factor measurements, in addition to further results arising from the hard scattering of polarized antiprotons and polarized protons at high $Q^{\,2}$ \cite{Barone:2005pu}.

The buildup of the polarization of an antiproton beam
by spin filtering off a polarized internal target (PIT)
has been described by differential equations
\cite{Milstein:2005bx,Nikolaev:2006gw,O'Brien:2007hu,MacKay:2006,Walcher:2007sj} involving the spin observables for antiproton-electron
and antiproton-proton elastic scattering, building on the earlier work of refs.~\cite{Krisch:1986nt,Meyer:1994,Horowitz:1994}.  This is of interest to the PAX collaboration who plan to polarize an antiproton beam by repeated interaction with a polarized internal target in a storage ring \cite{Barone:2005pu,Rathmann:2004pm}.  Relativistic expressions for spin observables due to single photon exchange in elastic spin 1/2 - spin 1/2  particle collisions have been presented recently \cite{O'Brien:2006zt}.  We use these spin observables to investigate the buildup of polarization of a beam of spin 1/2 particles by spin filtering.  

The paper is organized as follows: the method of polarization buildup by spin filtering is described in sect.~\ref{sec:Spin_filtering} and later sections each investigate a particular scenario of spin filtering.  Firstly spin filtering of a fully stored beam is described in sect.~\ref{sec:Polarization_buildup_of_a_stored_beam} and then spin filtering as the beam is being accumulated is described in sect.~\ref{sec:Accumulation_of_antiprotons_in_the_ring}.  In sect.~\ref{sec:Constant_beam_intensity} a system with constant beam intensity is analyzed, {\it i.e.}\ the input rate exactly balances the rate of loss of beam particles due to scattering out of the ring.  Polarization buildup of a stored beam which is initially polarized is the subject of sect.~\ref{sec:Stored_beam_with_initial_polarization}, and finally a system where particles are fed into the beam for a limited time after which spin filtering continues is treated in sect.~\ref{sec:Particles_fed_in_for_a_limited_time}.  The paper is summarized in sect.~\ref{sec:Summary}.

\section{Spin filtering}
\label{sec:Spin_filtering}

Spin filtering is the idea that a beam in a storage ring can be polarized by repeated interaction with a polarized internal target because of a difference between the cross-section for scattering of particles with their spins aligned parallel and the cross-section for scattering of particles with their spins aligned anti-parallel.  It was first proposed by P.~L.~Csonka in 1968 \cite{Csonka:1968}.  A ring acceptance angle $\theta_\mathrm{acc}$ divides the range of scattering angle $\theta$ into two distinct regions, a particle scattered an an angle below the acceptance angle remains {\bf ``in''} in beam and a particle scattered at an angle greater than the acceptance angle gets scattered {\bf ``out''} of the beam.  There is also a minimum scattering angle $\theta_{\mathrm{min}}$, corresponding to the average transverse electron separation for a pure electron target or to the Bohr radius of the atoms in an atomic target, below which scattering is prevented by Coulomb screening.  The two physical processes that contribute to polarization buildup in spin filtering are: (a) spin selective scattering out of the ring, and (b) selective spin-flip.  Thus particles in one spin state may be scattered out of the beam, or have their spin flipped while remaining in the beam, at a higher rate than particles in the other spin state.  Thus over time one spin state is depleted more than the other leading to a beam polarization.  

As the beam polarization increases the beam intensity decreases because particles are being continuously lost from the beam.  Thus there is a trade off between beam polarization and beam intensity that is characteristic of spin filtering.  The Figure Of Merit (FOM), defined as the polarization squared times the beam intensity $\mathrm{FOM}(\tau) = \mathcal{P}^{\,2}(\tau)\,N(\tau)$, where $\tau$ is the time elapsed since spin filtering began, provides a measure of the quality of the polarized beam taking this trade off into account.  We are investigating continuously inputing unpolarized particles into the beam, during spin filtering, to compensate the loss of beam intensity due to scattering out of the ring.

Note some treatments of spin filtering investigate a scenario where no particles are scattered out of the beam, {\it i.e.}\ the maximum scattering angle for the process is less than the ring acceptance angle, which is the case for antiprotons scattering off electrons in an atomic target \cite{Milstein:2005bx,Nikolaev:2006gw,O'Brien:2007hu,Meyer:1994,Horowitz:1994} and for antiprotons scattering off a co-moving beam of electrons or positrons \cite{Walcher:2007sj}.  In these scenarios only selective spin-flip can contribute to polarization buildup, and one avoids the problem of decreasing beam intensity.  The low density of the targets currently available causes the rate of polarization buildup using these methods to be slow, but the enhanced cross-sections at low energies suggested in refs.~\cite{Walcher:2007sj,Arenhovel:2007gi} may compensate this difficulty.  A comparison of all recent treatments of spin filtering, including a derivation of the polarization evolution equations treated here, is presented in ref.~\cite{O'Brien:2007jz}.

Spin filtering was demonstrated experimentally for polarized protons in 1993 by the FILTEX experiment at the Test Storage Ring in Heidelberg \cite{Rathmann:1993xf}.  The spin filtering method is at the heart of the recent PAX proposal to generate a polarized beam of antiprotons in the HESR ring of FAIR at GSI Darmstadt \cite{Barone:2005pu,Rathmann:2004pm}.

\section{Polarization buildup of a stored beam}
\label{sec:Polarization_buildup_of_a_stored_beam}

In this and the next few sections we describe systems of differential equations that model the buildup of polarization of a beam by spin filtering in a storage ring.  First a system where the beam has already been accumulated in the storage ring is analyzed.  The number of particles in the beam initially is $N_0$ and in this case the number of particles in the beam decreases continuously because of scattering out of the ring.  When circulating at frequency $\nu$, for a time $\tau$, in a ring with a polarized internal target of areal density $n$ and polarization $\mathcal{P}_T$ oriented normal to the ring plane, (or longitudinally with rotators)
\begin{eqnarray}
\label{eq:HomogeneousSystem}
  \frac{\mathrm{d}}{\mathrm{d}\tau}
\left[
        \begin{array}{c} N \\[2ex] J \end {array}
\right]
\,  = \, - \, n \, \nu
\left[
\begin{array}{ccc}
         I_\mathrm{\, out} && \mathcal{P}_T \, A_\mathrm{\, out}
\\[2ex]
    \mathcal{P}_T \, \left(\,A_\mathrm{\, all} -  K_\mathrm{\,in}\,\right)
&&
         I_\mathrm{\, all} -  D_\mathrm{\, in}
\end {array}
\right]
\,
\left[
        \begin{array}{c} N \\[2ex] J \end {array}
\right] \ ,
\end{eqnarray}
 describes the rate of change of the number of beam particles $N(\tau) = N_{\uparrow}(\tau) + N_{\downarrow}(\tau)$
 and their total spin $J(\tau) = N_{\uparrow}(\tau) - N_{\downarrow}(\tau)$ \cite{Milstein:2005bx,Nikolaev:2006gw}.
 These coupled differential equations involve angular integration of
 the spin observables presented in ref.~\cite{O'Brien:2006zt} over the following ranges, as seen in table~1.  The {\bf``in''} subscript refers to particles that are scattered at small angles $\leq \theta_{\mathrm{acc}}$ remaining in the beam, and the {\bf``out''} subscript refers to particles that are scattered out of the beam.  Thus the integrals over scattering angle $\theta$ are labeled {\bf``in''} where the range of integration is $\theta_{\mathrm{min}} \leq \theta \leq \theta_{\mathrm{acc}}$, {\bf``out''} where the range of integration is $\theta_{\mathrm{acc}} < \theta \leq \pi$ and {\bf``all''} $=$ {\bf``in''} $+$ {\bf``out''} where the range of integration is $\theta_{\mathrm{min}} \leq \theta \leq \pi$; as seen in table~1.  $I = {\mathrm{d}}\sigma\,/\,{\mathrm{d}}\Omega$ is the spin averaged differential cross-section and $A$, $K$ and $D$ are the double spin asymmetry, polarization transfer and depolarization spin observables respectively as calculated in ref.~\cite{O'Brien:2006zt}.  All cross-sections and spin observables contributing to spin filtering are azimuthally averaged, due to the geometry of the scattering, where the scattering plane can be at any azimuthal angle.  Hence single spin observables, for example the analyzing power, do not contribute to the polarization evolution equations because they vanish when azimuthally averaged.  The parameters in the matrix of coefficients of eq.~(\ref{eq:HomogeneousSystem}) depend on the state of the target polarization, {\it i.e.}\ longitudinal or transverse, as seen in table~1.
 
We now solve the set of polarization evolution equations presented in eq.~(\ref{eq:HomogeneousSystem}).  The eigenvalues of the matrix of coefficients are found to be 
\begin{eqnarray}
\lambda_1 & = & - \,n\,\nu\, \left(\,I_\mathrm{\,out} \,+\, L_\mathrm{\,in} \,+\, L_\mathrm{\,d}\,\right) \, ,\nonumber \\[2ex]
\lambda_2 & = & - \,n\,\nu\, \left(\,I_\mathrm{\,out} \,+\, L_\mathrm{\,in} \,-\, L_\mathrm{\,d}\,\right) \, ,
\end{eqnarray}
where the discriminant $L_\mathrm{\,d}$ of the quadratic equation
 for the eigenvalues is
\begin{equation}
   L_\mathrm{\,d}
\, =
\, \sqrt{\, \mathcal{P}_T^{\,2} \, A_\mathrm{\,out} \left( A_\mathrm{\,all}
\, -
\, K_\mathrm{\,in} \right) \, + \, L_\mathrm{\,in}^{\,2} }  \ \, ,
\end{equation}
and
$L_\mathrm{\,in} \, = \,\left( \, I_\mathrm{\,in} \, - \, D_\mathrm{\,in}\right)\,/\,2$
is a loss of polarization quantity.  Note that $I_\mathrm{\,out}$, $L_\mathrm{\,in}$ and $L_\mathrm{\,d}$ are all non-negative.  As a consequence the eigenvalues are non-positive and $\lambda_1 \leq \lambda_2 \leq 0$.  When there is no scattering out of the ring all of the {\bf \lq\lq out''} integrations are zero, and one finds $\lambda_1 \,=\, -\,2\,n\,\nu\,L_\mathrm{\,d}$ and $\lambda_2 \,=\, 0$.

Now enforcing the initial conditions $N(0) = N_0$ the total number of particles in the beam initially, and $J(0) = 0 \Rightarrow N_{\uparrow}(0) = N_{\downarrow}(0) = N_0\,/\,2$ {\it i.e.}\ initially the beam is unpolarized, gives the solutions:
\begin{eqnarray}
\label{eq:Homogeneous_N}
N(\tau) & = & \frac{\left[\,e^{\,\lambda_1\,\tau}\, \left(\,L_\mathrm{\,d} - L_\mathrm{\,in}\,\right) \, + \, e^{\,\lambda_2\,\tau}\, \left(L_\mathrm{\,d} + L_\mathrm{\,in}\right)\,\right]\,N_0}{2\, L_\mathrm{\,d}} \, ,\\[2ex]
\label{eq:Homogeneous_J}
J(\tau) & = & \displaystyle{ \frac{\left(\,e^{\,\lambda_1\,\tau}-e^{\,\lambda_2\,\tau}\,\right)\,\left(A_\mathrm{\,all} - K_\mathrm{\,in}\right)\,N_0\,\mathcal{P}_T}{2\, L_\mathrm{\,d}}}\,. 
\end{eqnarray}
The time ($\tau$) dependence of the polarization of the beam is given by
\begin{eqnarray}
\label{eq:HomogeneousPolarizationBuildup}
\mathcal{P}(\tau) \ = \ \frac{J(\tau)}{N(\tau)} \ = \ \frac{ -\,\left(\,A_\mathrm{\,all} \, - \, K_\mathrm{\,in}\,\right)\,\mathcal{P}_T}{L_\mathrm{\,in} \, + \, L_\mathrm{\,d} \, \coth\left(L_\mathrm{\,d}\, n\,\nu\,\tau\right)} \, .
\end{eqnarray}
The expression for $\mathcal{P}(\tau)$ is proportional to $\mathcal{P}_T$ which confirms that if the target polarization is zero there will be no polarization buildup in the beam, as was required by eqs.~(\ref{eq:HomogeneousSystem}).
The approximate rate of change of
 polarization for sufficiently short times, and the limit of the polarization for large times are respectively:
\begin{eqnarray}
\label{eq:Homogeneous_initial_and_max_polarization}
   \frac{\mathrm{d}\,\mathcal{P}}{\mathrm{d}\tau} & \approx & -\, n \, \nu \, \mathcal{P}_T
\,
\left( A_\mathrm{\,all} \, - \, K_\mathrm{\,in} \right) \, ,
\\[2ex]
\mathcal{P}_\mathrm{max} & = &  
   \displaystyle{\lim_{\tau \to \,\infty} \mathcal{P}(\tau) \, = -\, \mathcal{P}_T
\,
\frac{ A_\mathrm{\,all} \, - \, K_\mathrm{\,in}
}
{   L_\mathrm{\,in} \, + \, L_\mathrm{\,d}
}}
\,.
\end{eqnarray}
For pure electromagnetic scattering the double spin asymmetries equal the polarization transfer spin observables \cite{O'Brien:2006zt}, thus one can simplify the above equations using $A_\mathrm{\,in} = K_\mathrm{\,in}$, $A_\mathrm{\,out} = K_\mathrm{\,out}$ and $A_\mathrm{\,all} = K_\mathrm{\,all}$\,; hence $A_\mathrm{\,all} - K_\mathrm{\,in} = K_\mathrm{\,out}$\,.
\begin{table}[ht]
\label{tab:1}  
\begin{center}
\begin{tabular}{|c|c|} 
\hline
 & \\
Transverse polarization requires & Longitudinal polarization requires \\[2ex] \hline
 & \\
$\displaystyle{
\,
   I_\mathrm{out}
\, =
\, 2 \, \pi \!\int_{\theta_\mathrm{acc}}^{\pi}
\!
   \left(\frac{\mathrm{d}\,\sigma}{\mathrm{d}\,\Omega}\right) \sin\theta \, \mathrm{d}\theta
}
$
& 
$\displaystyle{
\,
   I_\mathrm{out}
\, =
\, 2 \, \pi \!\int_{\theta_\mathrm{acc}}^{\pi}
\!
   \left(\frac{\mathrm{d}\,\sigma}{\mathrm{d}\,\Omega}\right) \sin\theta \, \mathrm{d}\theta
}
$\\[3ex]
$\displaystyle{
    A_\mathrm{out}
   =
 2\,\pi \!\int_{\theta_\mathrm{acc}}^{\pi}
\!
   \left( \frac{A_\mathrm{XX} +  A_\mathrm{YY}}{2} \ 
   \frac{\mathrm{d}\,\sigma}{\mathrm{d}\,\Omega} \right) \sin\theta \, \mathrm{d}\theta 
}
$
&
$\displaystyle{
\,
    A_\mathrm{out}
\,
   =
\, 2\,\pi \!\int_{\theta_\mathrm{acc}}^{\pi}
\!
   \left( A_\mathrm{ZZ} \ 
   \frac{\mathrm{d}\,\sigma}{\mathrm{d}\,\Omega} \right) \sin\theta \, \mathrm{d}\theta 
}
$\\[3ex]
$ \displaystyle{
   A_\mathrm{all}
 =
 2\,\pi \!\int_{\theta_\mathrm{min}}^{\pi}
\!
  \left( \frac{A_\mathrm{XX}  +  A_\mathrm{YY}}{2} \ 
   \frac{\mathrm{d}\,\sigma}{\mathrm{d}\,\Omega} \right) \sin\theta \, \mathrm{d}\theta
}
$
&
$ \displaystyle{ 
\,  
   A_\mathrm{all}
\, =
\, 2\,\pi \!\int_{\theta_\mathrm{min}}^{\pi}
\!
    \left( A_\mathrm{ZZ} \ 
   \frac{\mathrm{d}\,\sigma}{\mathrm{d}\,\Omega} \right) \sin\theta \, \mathrm{d}\theta
}
$\\[3ex]
$\displaystyle{
    K_\mathrm{in}
   =
  2\,\pi \!\int_{\theta_\mathrm{min}}^{\theta_\mathrm{acc}}
\!
   \left( \frac{K_\mathrm{XX} +  K_\mathrm{YY}}{2} \ 
   \frac{\mathrm{d}\,\sigma}{\mathrm{d}\,\Omega} \right)  \sin\theta \, \mathrm{d}\theta
}
$
&
$\displaystyle{
\,
    K_\mathrm{in}
\,
   =
\,  2\,\pi \!\int_{\theta_\mathrm{min}}^{\theta_\mathrm{acc}}
\!
  \left( K_\mathrm{ZZ}\ 
   \frac{\mathrm{d}\,\sigma}{\mathrm{d}\,\Omega} \right) \sin\theta \, \mathrm{d}\theta
}
$\\[3ex]
$\displaystyle{
    D_\mathrm{in}
   =
  2\,\pi \!\int_{\theta_\mathrm{min}}^{\theta_\mathrm{acc}}
\!
   \left( \frac{D_\mathrm{XX}  +  D_\mathrm{YY}}{2}\ 
   \frac{\mathrm{d}\,\sigma}{\mathrm{d}\,\Omega} \right)  \sin\theta \, \mathrm{d}\theta
}
$
&
$\displaystyle{
\,
    D_\mathrm{in}
\,
   =
\,  2\,\pi \!\int_{\theta_\mathrm{min}}^{\theta_\mathrm{acc}}
\!
  \left( D_\mathrm{ZZ}\ 
   \frac{\mathrm{d}\,\sigma}{\mathrm{d}\,\Omega} \right)  \sin\theta \, \mathrm{d}\theta
}$
\\[3ex]\hline
\end{tabular} 
\caption{The entries in the system of equations for polarization buildup involve angular integration over the spin observables presented in refs.~\cite{O'Brien:2006zt,Arenhovel:2007gi,Leader:2005}.  $\mathrm{X}$, $\mathrm{Y}$ and $\mathrm{Z}$ are the coordinate axis where the beam is moving in the positive $\mathrm{Z}$ direction.  The minimum value for $\theta$ ($\theta_\mathrm{min}$) relates to the average transverse electron separation for a pure electron target and to the Bohr radius for an atomic gas target, and $\theta_\mathrm{acc}$ is the ring acceptance angle.}
\end{center}
\end{table}

Eqs.~(\ref{eq:HomogeneousSystem}) neglect a spin tune effect \cite{Nikolaev:2006gw}, induced by a pseudomagnetic field generated by the target polarization dependent real part of the forward scattering amplitude.  The effect which is expected to be comparatively small comes from the difference between the spin elastic helicity amplitudes, which are known for electromagnetic scattering from QED \cite{O'Brien:2006zt,Buttimore:1978ry}, but not well known for hadronic scattering even though they could be evaluated from dispersion relations for the appropriate spin dependent total cross-sections.

\subsection{Beam lifetime and figure of merit}
\label{sec:Beam_lifetime_and_figure_of_merit}

The beam lifetime $\tau_*$, the time taken for the beam intensity to decrease by a factor of $e$, can be obtained from eq.~(\ref{eq:Homogeneous_N}).  One finds 
\begin{equation}
\label{eq:Beam_lifetime}
\tau_{*} \ \approx \ \displaystyle{\frac{1}{n\,\nu\,\left(\,I_\mathrm{\,out} + L_\mathrm{\,in} \,\right)}} \, ,
\end{equation}
 where $I_\mathrm{\,out} + L_\mathrm{\,in} \gg L_\mathrm{\,d}$ when there is scattering out of the ring.

The Figure Of Merit (FOM) provides a measure of the quality of the polarized beam, and is given by
\begin{equation}
\mathrm{FOM}(\tau) \ = \ \mathcal{P}^{\,2}(\tau)\ N(\tau) \ = 
 \ \frac{J^{\,2}(\tau)}{N(\tau)} \,.
\end{equation}
The figure of merit for the above case is
\begin{eqnarray}
\label{eq:Homogeneous_FOM}
\hspace*{-3em}\mathrm{FOM}(\tau) & = & \frac{\left(A_\mathrm{\,all} - K_\mathrm{\,in}\right)^{\,2} N_0\,\mathcal{P}_T^{\,2}}{2\,L_\mathrm{\,d}}\,\left[ \,\frac{\displaystyle{\left(\,e^{\,\lambda_1\,\tau} - e^{\,\lambda_2\,\tau}\,\right)^{\,2}}}{e^{\,\lambda_1\,\tau}\, \left(\,L_\mathrm{\,d} - L_\mathrm{\,in}\,\right) \, + \, e^{\,\lambda_2\,\tau}\, \left(L_\mathrm{\,d} + L_\mathrm{\,in}\right)}\,\right] \, .
\end{eqnarray}
Maximizing the figure of merit gives the optimum polarization buildup time, taking into account the trade-off between decreasing beam intensity and increasing beam polarization.  Solving $\mathrm{d}\,\mathrm{FOM}\,/\,\mathrm{d}\,\tau \, = \, 0$ yields $\tau_\mathrm{optimum} \, \approx \, 2 \,/ \,n\,\nu\,\left(\,I_\mathrm{\,out} + L_\mathrm{\,in}\,\right)$, approximately twice the beam lifetime.  Thus the optimum time for polarization buildup is twice the lifetime of the beam, as mentioned in ref.~\cite{Rathmann:2004pm}.

\section{Accumulation of antiprotons in the ring}
\label{sec:Accumulation_of_antiprotons_in_the_ring}

In the discussion so far we have only considered polarizing an antiproton beam when the beam is already accumulated in the storage ring.  The PAX collaboration plans to obtain their antiproton beam by collecting the produced antiprotons from high energy interactions of protons on targets of light nuclei, such as Beryllium.  The antiprotons will be continuously fed into the storage ring at a fixed rate and accumulated, hence increasing the beam intensity, allowing for a greater luminosity in an experiment.  The PAX collaboration estimates the production rate of antiprotons as being $10^7$ per second \cite{Barone:2005pu}.  Since $10^{11}$ antiprotons are required in the storage ring, antiprotons will be fed into the storage ring at a rate of $10^7$ per second for $10^4$ seconds \cite{Barone:2005pu}.

We now consider a system where spin filtering occurs as the antiprotons are being fed into the ring.  The original system of equations must be amended to account for this constant accumulation.  The effect will be to add a term $\beta$ to the $\mathrm{d}\,N(\tau)\,/\,\mathrm{d}\,\tau$ equation, where $\beta$ is the constant rate at which antiprotons are fed into the ring; while the $\mathrm{d}\,J(\tau)\,/\,\mathrm{d}\,\tau$ equation remains unchanged.  The initial conditions are $N(0) = N_0$\,, which will be set to zero in sect.~\ref{sec:No_initial_beam}, and $J(0) = 0$.  The new system of differential equations is
\begin{eqnarray}
\label{eq:InhomogeneousSystem_N}
\frac{\mathrm{d}\,N(\tau)}{\mathrm{d}\,\tau} & = & \, - \, n \, \nu \ \left[\, I_\mathrm{\, out} \ N(\tau) \ + \  \mathcal{P}_T \, A_\mathrm{\, out} \ J(\tau)\, \right] \ + \  \beta  \, , \\[1ex]
\label{eq:InhomogeneousSystem_J}
\frac{\mathrm{d}\,J(\tau)}{\mathrm{d}\,\tau} & = &\, - \, n \, \nu \  \left[\, \mathcal{P}_T \, \left(\,A_\mathrm{\, all} -  K_\mathrm{\,in}\,\right)\ N(\tau) \  + \ \left(\,I_\mathrm{\, all} -  D_\mathrm{\, in}\,\right)\ J(\tau)\, \right] \,.
\end{eqnarray}
By differentiating eq.~(\ref{eq:InhomogeneousSystem_J}) with respect to $\tau$ and substituting in eq.~(\ref{eq:InhomogeneousSystem_N}) one obtains an inhomogeneous second order linear differential equation with constant coefficients for $J(\tau)$:
\begin{equation}
\label{eq:Inhomogeneous_Second_Order_ODE}
\frac{\mathrm{d}^{\,2}\,J(\tau)}{\mathrm{d}\,\tau^2} \ - \ \left(\,\lambda_1 + \lambda_2\,\right)\,\frac{\mathrm{d}\,J(\tau)}{\mathrm{d}\,\tau} \ +\  \lambda_1\,\lambda_2\,J(\tau) \ = \  - \,n\, \nu \,\mathcal{P}_T\,\left(\,A_\mathrm{\,all} - K_\mathrm{\,in}\,\right) \, \beta \,,
\end{equation}
the solution of which is
\begin{eqnarray}
\label{eq:Inhomogeneous_J}
J(\tau) & = & \frac{\mathcal{P}_T\,\left(\,A_\mathrm{all} - K_\mathrm{in}\,\right)}{2\,L_\mathrm{d}\,\lambda_1\,\lambda_2}\ \left[\,\lambda_2\,\left(\,\lambda_1\,N_0 + \beta \,\right)\,e^{\,\lambda_1\,\tau} \ - \ \lambda_1\,\left(\,\lambda_2\,N_0 + \beta \,\right)\,e^{\,\lambda_2\,\tau}  \right. \nonumber \\[1ex]
& & \qquad \qquad \qquad \qquad \left. \ + \ \beta\,\left(\,\lambda_1 - \lambda_2\,\right) 
\,\right] \,.
\end{eqnarray}
Differentiating eq.~(\ref{eq:Inhomogeneous_J}) with respect to $\tau$ and 
substituting into eq.~(\ref{eq:InhomogeneousSystem_J}) gives an expression for $N(\tau)$\,:
\begin{eqnarray}
\label{eq:Inhomogeneous_N}
N(\tau)  & = & \frac{1}{2\,L_\mathrm{d}\,\lambda_1\,\lambda_2} \left[\,
\lambda_2\,\left(\,\lambda_1\,N_0 + \beta \,\right) \,\left(\,L_\mathrm{d} - L_\mathrm{in}\,\right)\,e^{\,\lambda_1\,\tau}
 \right. \\[2ex]
& & \left.+ \, \lambda_1\,\left(\,\lambda_2\,N_0 + \beta\,\right) \,\left(\,L_\mathrm{in} + L_\mathrm{d}\,\right)\,e^{\,\lambda_2\,\tau} \, + \, \beta\,\left(\,I_\mathrm{all} - D_\mathrm{in}\,\right)\,\left(\,\lambda_2 - \lambda_1\,\right)\,\right]\nonumber \,.
\end{eqnarray}
As a consistency check one can see that these solutions for $J(\tau)$ and $N(\tau)$ satisfy the initial conditions $J(0) = 0$ and $N(0) = N_0$, and in the particular case when $\beta = 0$ the above expressions reduce to the solution of the homogeneous system presented in sect.~\ref{sec:Polarization_buildup_of_a_stored_beam}.
Dividing $J(\tau)$ by $N(\tau)$ we obtain an expression for the polarization $\mathcal{P}(\tau)$ as a function of time,
\begin{equation}
\label{eq:Inhomogeneous_Polarization_Buildup}
\mathcal{P}(\tau) \ = \ \frac{-\,\left(\,A_\mathrm{\,all} - K_\mathrm{\,in}\,\right)\,\mathcal{P}_T}{L_\mathrm{\,in} + L_\mathrm{\,d}\,\displaystyle{\left[\,\frac{2}{1 - \frac{\lambda_2\,\left[\,e^{\,\lambda_1\,\tau}\,\left(\,\lambda_1\,N_0 + \beta\,\right)\, - \, \beta\,\right]
}{\lambda_1\,\left[\,e^{\,\lambda_2\,\tau}\,\left(\,\lambda_2\,N_0 + \beta\,\right)\, - \, \beta\,\right]}}\,-\,1\,\right]
}}\,.
\end{equation}
When the particle input rate is zero ({\it i.e.}\ $\beta = 0$) the above equation simplifies to 
\begin{equation}
\mathcal{P}(\tau) \, = \, \frac{-\,\left(\,A_\mathrm{\,all} - K_\mathrm{\,in}\,\right)\,\mathcal{P}_T}{L_\mathrm{\,in} + L_\mathrm{\,d}\,\left[\,\displaystyle{\frac{2}{1 - e^{\,\left(\,\lambda_1 - \lambda_2\,\right)\,\tau}}} - 1 \,\right]}
\, = \, 
\frac{ -\,\left(\,A_\mathrm{\,all} \, - \, K_\mathrm{\,in}\,\right)\,\mathcal{P}_T}{L_\mathrm{\,in} \, + \, L_\mathrm{\,d} \, \coth\left(L_\mathrm{\,d}\, n\,\nu\,\tau\right)} \, ,
\end{equation}
which is the solution of the homogeneous case presented in eq.~(\ref{eq:HomogeneousPolarizationBuildup}).

Using a Taylor Series expansion we find that the approximate initial rate of polarization buildup for each of these cases ($N_0 \neq 0$ with $\beta \neq 0$ and $N_0 = 0$ with $\beta \neq 0$) is the same as in the homogeneous case ($N_0 \neq 0$ with $\beta = 0$):
\begin{equation}
   \frac{\mathrm{d}\,\mathcal{P}}{\mathrm{d}\tau} \, \, \approx \, -\, n \, \nu \, \mathcal{P}_T
\,
\left( A_\mathrm{\,all} \, - \, K_\mathrm{\,in} \right)\,.
\end{equation}
The maximum polarization achievable is the limit as time approaches infinity:
\begin{equation}
\mathcal{P}_\mathrm{max} \, = \,  \lim_{\tau \to \,\infty} \mathcal{P}(\tau) \, = \, \frac{-\,\mathcal{P}_T\,\left(\,A_\mathrm{\,all} - K_\mathrm{\,in}\,\right)}{I_\mathrm{\,all} - D_\mathrm{\,in}} \ ,
\end{equation}
which is independent of both $N_0$ and $\beta$, however note that in taking this limit we used the fact that $\beta \neq 0$.  If $\beta$ was equal to zero then the maximum polarization achievable would equal that from the homogeneous case; as can be easily seen from eq.~(\ref{eq:Inhomogeneous_Polarization_Buildup}) remembering that $\lambda_1 \leq \lambda_2 \leq 0$.  Thus for the complete case there are just two values of the maximum polarization, one for $\beta = 0$ and one for all $\beta \neq 0$.
The figure of merit for this inhomogeneous case is:
\begin{eqnarray}
\mathrm{FOM}(\tau) & = & \mathcal{P}^{\,2}(\tau)\,N(\tau) \ = \ \frac{J^{\,2}(\tau)}{N(\tau)} \ = \ \frac{\left(\,A_\mathrm{\,all} - K_\mathrm{\,in}\,\right)^{\,2}\,\mathcal{P}_T^{\,2}}{2\,L_\mathrm{\,d}\,\lambda_1\,\lambda_2} \times \nonumber\\[2ex]
& & \hspace*{-8em}   \frac{\left[\,
c_1\,e^{\,\lambda_1\,\tau} - c_2\,e^{\,\lambda_2\,\tau} + \beta\,\left(\,\lambda_1 - \lambda_2\,\right)\,\right]^{\,2}}{c_1\,\left(\,L_\mathrm{\,d} - L_\mathrm{\,in}\,\right)\,e^{\,\lambda_1\,\tau} +  c_2\,\left(\,L_\mathrm{\,in} + L_\mathrm{\,d}\,\right)\,e^{\,\lambda_2\,\tau} + \beta\,\left(\,I_\mathrm{\,all} - D_\mathrm{\,in}\,\right)\,\left(\,\lambda_2 - \lambda_1\,\right)} \, ,
\end{eqnarray}
where for convenience we have defined the two constants $c_1 = \lambda_2\,\left(\,\lambda_1\,N_0 + \beta\,\right)$ and $c_2 = \lambda_1\,\left(\,\lambda_2\,N_0 + \beta\,\right)$.  Note the FOM will not have a maximum in finite time if the accumulation rate $\beta$ is high enough to make the beam intensity a constant or increase with time.  If this happens the FOM will increase monotonically.

\subsection{No initial beam}
\label{sec:No_initial_beam}

Of interest is the particular case when $N_0 = 0$, {\it i.e.}\ there are no particles in the beam initially.  In this case the above solutions simplify to
\begin{eqnarray}
\label{eq:Inhomogeneous_J_with_N0_0}
J(\tau) & = & \frac{\beta\,\mathcal{P}_T\,\left(\,A_\mathrm{all} - K_\mathrm{in}\,\right)}{2\,L_\mathrm{d}\,\lambda_1\,\lambda_2}\ \left[\, \lambda_2\,\left(\,e^{\,\lambda_1\,\tau} - 1\,\right) \,+ \, \lambda_1\, \left(\,1-e^{\,\lambda_2\,\tau}\,\right)\,\right]  \, ,\\[2ex]
\label{eq:Inhomogeneous_N_with_N0_0}
N(\tau)  & = & \frac{\beta}{2\,L_\mathrm{d}\,\lambda_1\,\lambda_2} \left[\,
\lambda_2\,\left(\,L_\mathrm{d} - L_\mathrm{in}\,\right)\,e^{\,\lambda_1\,\tau}
\,+ \, \lambda_1\,\left(\,L_\mathrm{in} + L_\mathrm{d}\,\right)\,e^{\,\lambda_2\,\tau} \right. \\[2ex]
& & \left. \qquad \qquad \qquad + \ \left(\,I_\mathrm{all} - D_\mathrm{in}\,\right)\,\left(\,\lambda_2 - \lambda_1\,\right)\,\right] \, ,\nonumber \\[2ex]
\label{eq:Polarization_Inhomogeneous_N_equals_0}
\mathcal{P}(\tau) &  = &  \frac{-\,\left(\,A_\mathrm{\,all} - K_\mathrm{\,in}\,\right)\,\mathcal{P}_T}{L_\mathrm{\,in} + \displaystyle{L_\mathrm{\,d}\, \left[\,\frac{2}{1 - \frac{\left(\,1 - e^{\,\lambda_1\,\tau}\,\right)\,\lambda_2}{\left(\,1 - e^{\,\lambda_2\,\tau}\,\right)\,\lambda_1}}\,-\,1\,\right] 
}}\,.
\end{eqnarray}
Interestingly the $\beta$ dependence of $\mathcal{P}(\tau)$ vanishes in this case, {\it i.e.}\ the polarization buildup rate is independent of the rate at which antiprotons are fed into the ring, if there are no particles in the beam initially.  But we have used the fact that $\beta \neq 0$ to obtain the above result.   We should note the obvious physical fact that if $N_0 = 0$ and $\beta = 0$ {\it i.e.}\ there are no particles in the beam initially and no particles are fed into the beam, then there will never be any particles in the beam; so measuring the beam polarization is meaningless.  The accumulation rate $\beta$ will effect the figure of merit as it greatly effects the beam intensity $N(\tau)$.

We can summarize the results for the polarization buildup as
\begin{eqnarray}
\mathcal{P}(\tau) \ = \ \left\{
\begin{array}{ll}
0 & \mbox{for} \ \beta = 0 \ \ \& \ \ N_0 = 0\\[5ex]
\displaystyle{\frac{ -\,\left(\,A_\mathrm{\,all} \, - \, K_\mathrm{\,in}\,\right)\,\mathcal{P}_T}{L_\mathrm{\,in} \, + \, L_\mathrm{\,d} \, \coth\left(L_\mathrm{\,d}\, n\,\nu\,\tau\right)}}  & \mbox{for} \ \beta = 0 \ \ \& \ \ N_0 \neq 0 \nonumber\\[5ex]
\displaystyle{\frac{-\,\left(\,A_\mathrm{\,all} - K_\mathrm{\,in}\,\right)\,\mathcal{P}_T}{L_\mathrm{\,in} + L_\mathrm{\,d}\,\displaystyle{\left[\,\frac{2}{1 - \frac{\left(\,1 - e^{\,\lambda_1\,\tau}\,\right)\,\lambda_2}{\left(\,1 - e^{\,\lambda_2\,\tau}\,\right)\,\lambda_1}}\,-\,1\,\right]
}}} &  \mbox{for} \ \beta \neq 0 \ \ \& \ \ N_0 = 0 \nonumber \\[13ex]
\displaystyle{
\frac{-\,\left(\,A_\mathrm{\,all} - K_\mathrm{\,in}\,\right)\,\mathcal{P}_T}{L_\mathrm{\,in} + L_\mathrm{\,d}\,\displaystyle{\left[\,\frac{2}{1 - \frac{\lambda_2\,\left[\,e^{\,\lambda_1\,\tau}\,\left(\,\lambda_1\,N_0 + \beta\,\right)\, - \, \beta\,\right]
}{\lambda_1\,\left[\,e^{\,\lambda_2\,\tau}\,\left(\,\lambda_2\,N_0 + \beta\,\right)\, - \, \beta\,\right]}}\,-\,1\,\right]
}}} &  \mbox{for} \ \beta \neq 0 \ \ \& \ \ N_0 \neq 0 \nonumber 
\end {array}
\right.\nonumber
\end{eqnarray}

\section{Constant beam intensity}
\label{sec:Constant_beam_intensity}

In this case the accumulation rate is set specifically so that extra particles are fed into the beam at such a rate so that the beam intensity is kept constant, {\it i.e.}\ fed in at such a rate to balance the rate at which particles are scattered out of the beam.  The system of equations is much simpler in this case.   Here $N(\tau) = N_0$ is a constant, hence $\mathrm{d}\,N(\tau)/\mathrm{d}\,\tau =0$, and the $J(\tau)$ equation becomes a 1st order linear ODE with constant coefficients
\begin{equation}
\frac{\mathrm{d}\,J(\tau)}{\mathrm{d}\,\tau} \, + \,  n\,\nu \, \left(\,I_\mathrm{\,all} - D_\mathrm{\,in}\,\right)\,J(\tau) \, = \,  - \, n\,\nu \, \left(\,A_\mathrm{\,all} - K_\mathrm{\,in}\,\right)\,\mathcal{P}_T \ N_0 \,,
\end{equation}
and imposing the initial conditions $N(0) = N_0$ and $J(0) = 0$ one obtains the solution
\begin{eqnarray}
J(\tau) & = & \frac{\left(\,A_\mathrm{\,all} - K_\mathrm{\,in}\,\right)\,\mathcal{P}_T \ N_0}{\left(\,I_\mathrm{\,all} - D_\mathrm{\,in}\,\right)} \ \left[\,e^{\,-\, n\,\nu \, \left(\,I_\mathrm{\,all} \ - \ D_\mathrm{\,in}\,\right)\ \tau} -1\,\right] \,.
\end{eqnarray}
Now the polarization as a function of time can be presented
\begin{equation}
\mathcal{P}(\tau) \, = \, \frac{J(\tau)}{N(\tau)} \, = \, \frac{J(\tau)}{N_0} \, = \ \frac{\left(\,A_\mathrm{\,all} - K_\mathrm{\,in}\,\right)\,\mathcal{P}_T}{\left(\,I_\mathrm{\,all} - D_\mathrm{\,in}\,\right)} \ \left[\,e^{\,-\, n\ \nu \  \left(\,I_\mathrm{\,all}\ - \ D_\mathrm{\,in}\,\right)\ \tau} -1\,\right] \,.
\end{equation}
To find the maximum polarization achievable, {\it i.e.}\ the limit as time tends to infinity, we note that $I_\mathrm{\,all} > D_\mathrm{\,in}$ thus $-\, n\,\nu \, \left(\,I_\mathrm{\,all} - D_\mathrm{\,in}\,\right) < 0$ and hence one obtains
\begin{equation}
\mathcal{P}_\mathrm{max} \ = \ \lim_{\tau \to \,\infty} \mathcal{P}(\tau) \ = \ \frac{-\,\mathcal{P}_T\,\left(\,A_\mathrm{\,all} - K_\mathrm{\,in}\,\right)}{I_\mathrm{\,all} - D_\mathrm{\,in}} \,,
\end{equation}
which is the same as in the inhomogeneous case when $\beta \neq 0$.

The initial rate of polarization buildup can be obtained by expanding $\mathcal{P}(\tau)$ as a Taylor expansion in $n\,\nu\,\tau$.  Assuming $n\,\nu\,\tau$ is small we neglect terms of second or higher order giving  
\begin{equation}
\frac{\mathrm{d}\,\mathcal{P}}{\mathrm{d}\,\tau} \  \approx \   -\,n\,\nu\,\left(\,A_\mathrm{\,all} - K_\mathrm{\,in}\,\right)\,\mathcal{P}_T \, ,
\end{equation}
as it was in the homogeneous case presented in eq.~(\ref{eq:Homogeneous_initial_and_max_polarization}).  The figure of merit in this case is easily obtained
\begin{eqnarray}
\mathrm{FOM}(\tau) 
& = & \frac{N_0\,\mathcal{P}_T^{\,2}\ \left(\,A_\mathrm{\,all} - K_\mathrm{\,in}\,\right)^{\,2}}{\left(\,I_\mathrm{\,all} - D_\mathrm{\,in}\,\right)^{\,2}} \ \left[\,1 \, - \, e^{\,-\, n\ \nu \  \left(\,I_\mathrm{\,all}\ - \ D_\mathrm{\,in}\,\right)\ \tau}\,\right]^{\,2} \,.
\end{eqnarray}
and increases monotonically with time.

\subsection{Approximating the critical input rate}
\label{subsec:Approximating_the_critical_input_rate}

The accumulation rate needed to keep the beam intensity constant is important, as this critical rate divides the solution of the system into two physically distinct cases.  Smaller accumulation rates than this critical value cause the beam intensity to decrease, hence the FOM will have a maximum in finite time.  Larger values than the critical value cause the beam intensity to increase continuously, hence the FOM will increase monotonically.  We can see from eq.~(\ref{eq:Homogeneous_N}) that $N(\tau)$ does not decrease linearly with time $\tau$.  So the accumulation rate needed to keep the beam intensity constant, say $f(\tau)$, will not be linear in $\tau$.  We now derive the function  $f(\tau)$ and obtain a linear approximation to it, which can be used in the inhomogeneous case treated in sect.~\ref{sec:Accumulation_of_antiprotons_in_the_ring}.  We must solve $N(\tau) = N_{hom}(\tau) + f(\tau) = N_0$, {\it i.e.}\
\begin{equation}
N(\tau)  \, = \,  \frac{\left[\,e^{\,\lambda_1\,\tau}\, \left(\,L_\mathrm{\,d} - L_\mathrm{\,in}\,\right) \, + \, e^{\,\lambda_2\,\tau}\, \left(L_\mathrm{\,d} + L_\mathrm{\,in}\right)\,\right]\,N_0}{2\, L_\mathrm{\,d}} + f(\tau) \, = \, N_0 \, ,
\end{equation}
which leads to
\begin{eqnarray}
f(\tau) 
 & = & \frac{N_0}{2\, L_\mathrm{\,d}}\ \left[\,2\, L_\mathrm{\,d} - e^{\,\lambda_1\,\tau}\, \left(\,L_\mathrm{\,d} - L_\mathrm{\,in}\,\right) - e^{\,\lambda_2\,\tau}\, \left(L_\mathrm{\,d} + L_\mathrm{\,in}\right)\,\right] \,.
\end{eqnarray}
A linear approximation $f_L(\tau)$ to $f(\tau)$ can be found by Taylor expanding the exponentials to order $\tau$ (which is valid since $n\,\nu\,\tau$ is small), to obtain
\begin{equation}
f_L(\tau)  \ = \ n\,\nu\,N_0\, I_\mathrm{\,out}\,\tau \, ,
\end{equation}
which is in the linear form $+\,\beta_c\,\tau$ where $\beta_c = n\,\nu\,N_0\, I_\mathrm{\,out}$ is the critical value of $\beta$ which when added to the $\mathrm{d}\,N(\tau)\,/\,\mathrm{d}\,\tau$ differential equation  in sect.~\ref{sec:Accumulation_of_antiprotons_in_the_ring} approximately makes the beam intensity constant.

\section{Stored beam with initial polarization}
\label{sec:Stored_beam_with_initial_polarization}

We now solve the homogeneous system where the initial polarization is not zero.  This will be used if two methods of polarizing antiprotons are combined, {\it i.e.}\ if antiprotons were produced with a small polarization by some other method and one wanted to increase that polarization by spin filtering in a storage ring, where the luminosity could also be increased.    In this section the beam has been stored and there is no further input of particles into the beam.

The system of differential equations, eigenvalues and eigenvectors are the same as sect.~\ref{sec:Polarization_buildup_of_a_stored_beam}, but one of the initial conditions is different.  The new initial conditions are $N(0) = N_0 > 0$ the total number of particles in the beam initially, and $J(0) = J_0 \neq 0 \Rightarrow N_{\uparrow}(0) \neq N_{\downarrow}(0)$ {\it i.e.}\ initially the beam is polarized.  Note that since the number of particles in the beam with one particular spin state must not be greater than the total number of particles in the beam the bound $|J_0| \leq N_0$ is respected.  A negative value for $J_0$ simply implies that the antiproton beam is initially polarized in the opposite direction to the polarization direction of the target.  
Enforcing these initial conditions leads to the solutions
\begin{equation}
\label{eq:Initial_Polarization_N}
 N(\tau) \ = \  \frac{\left(\,J_0\,\mathcal{P}_T\,A_\mathrm{\,out} \,-\,N_0\,L_\mathrm{\, in}\,\right)\, \left(\,e^{\,\lambda_1\,\tau}\, - \, e^{\,\lambda_2\,\tau}\,\right)\, +\, N_0 \,L_\mathrm{\, d}\, \left(\,e^{\,\lambda_1\,\tau} \,+\, e^{\,\lambda_2\,\tau}\,\right)}{2\,L_\mathrm{\, d}}
\end{equation}
\vspace*{1ex}
\begin{equation}
\label{eq:Initial_Polarization_J}
\hspace*{-0.5em} J(\tau) \, = \, \frac{\left[\,N_0 \,\mathcal{P}_T \left(\,A_\mathrm{\,all} - K_\mathrm{\,in}\,\right) + J_0\,L_\mathrm{\, in} \,\right] \left(\,e^{\,\lambda_1\,\tau} - e^{\,\lambda_2\,\tau}\,\right) + J_0\,L_\mathrm{\, d} \left(\,e^{\,\lambda_1\,\tau} + e^{\,\lambda_2\,\tau}\,\right)}{2\,L_\mathrm{\, d}}
\end{equation}
which reduce to the solutions of the original homogeneous system presented in eqs.~(\ref{eq:Homogeneous_N},~\ref{eq:Homogeneous_J}) when $J_0 \rightarrow 0$.  The beam lifetime is the same to leading approximation as in the homogeneous case when $J(0) = 0$.  Dividing $J(\tau)$ by $N(\tau)$ provides the time dependence of the polarization of the beam
\begin{eqnarray}
\mathcal{P}(\tau)  & = &   \frac{L_\mathrm{\, d}\,J_0 - \,\tanh\left(L_\mathrm{\, d}\,n\,\nu\,\tau\,\right)\,\left[\,L_\mathrm{\, in}\,J_0 + N_0\,\mathcal{P}_T\,\left(\,A_\mathrm{\,all} - K_\mathrm{\,in}\,\right)\,\right]}{L_\mathrm{\, d}\,N_0 + \tanh\left(L_\mathrm{\, d}\,n\,\nu\,\tau\,\right)\,\left[\,L_\mathrm{\, in}\,N_0 - J_0\,\mathcal{P}_T\,A_\mathrm{\,out}\,\right]} \,.
\end{eqnarray}
Denoting the initial polarization $\mathcal{P}(0) = J_0\,/\,N_0 \equiv \mathcal{P}_0$ the above can be written as
\begin{equation}
\label{eq:Polarization_buildup_with_initial_polarization}
\mathcal{P}(\tau)  \ = \  \frac{L_\mathrm{\, d}\,\mathcal{P}_0 - \tanh\left(L_\mathrm{\, d}\,n\,\nu\,\tau\,\right)\,\left[\,L_\mathrm{\, in}\,\mathcal{P}_0 + \mathcal{P}_T\,\left(\,A_\mathrm{\,all} - K_\mathrm{\,in}\,\right)\,\right]}{L_\mathrm{\, d} + \tanh\left(L_\mathrm{\, d}\,n\,\nu\,\tau\,\right)\,\left[\,L_\mathrm{\, in}\, - \mathcal{P}_0\,\mathcal{P}_T\,A_\mathrm{\,out}\,\right]}\,.
\end{equation}
The approximate rate of change of polarization for sufficiently short times is found by Taylor expanding to first order in $\tau$
\begin{equation}
   \frac{\mathrm{d}\,\mathcal{P}}{\mathrm{d}\tau} \, \ \approx \  n\,\nu\,\left\{\,\left[\,A_\mathrm{\,out}\,\mathcal{P}_0^{\,2} - \left(\,A_\mathrm{\,all} - K_\mathrm{\,in}\,\right)\,\right]\,\mathcal{P}_T\, - 2\,\mathcal{P}_0\,L_\mathrm{\, in}\,\right\}\,.
\end{equation}
The limit as time goes to infinity of $\mathcal{P}(\tau)$ in eq.~(\ref{eq:Polarization_buildup_with_initial_polarization}) is
\begin{equation}
 \lim_{\tau \to \,\infty} \ \mathcal{P}(\tau) \ = 
\ 
\frac{ \mathcal{P}_0\,\left(\,L_\mathrm{\,d} \, - \, L_\mathrm{\,in}\,\right)\,-\, \mathcal{P}_T\,\left(\,A_\mathrm{\,all} \, - \, K_\mathrm{\,in}\,\right)
}
{   \left(\,L_\mathrm{\,in} \, + \, L_\mathrm{\,d}\,\right) - A_\mathrm{\,out}\,\mathcal{P}_0\ \mathcal{P}_T
} \, ,
\end{equation}
which of course agrees with the earlier maximum polarization if $J_0 = 0$ ({\it i.e.}\ $\mathcal{P}_0 = 0$).  
The figure of merit for this case is:
\begin{eqnarray}
\mathrm{FOM}(\tau) & = & \mathcal{P}^{\,2}(\tau)\,N(\tau) \ = \ \frac{J^{\,2}(\tau)}{N(\tau)} \ = \\[2ex]
& & \hspace*{-6em}  \displaystyle{ \frac{N_0\,\left\{\,\left[\,\mathcal{P}_T\,\left(\,A_\mathrm{\,all} - K_\mathrm{\,in}\,\right) + \mathcal{P}_0\,L_\mathrm{\, in} \,\right] \left(\,e^{\,\lambda_1\,\tau} - e^{\,\lambda_2\,\tau}\,\right) + \mathcal{P}_0\,L_\mathrm{\, d} \left(\,e^{\,\lambda_1\,\tau} + e^{\,\lambda_2\,\tau}\,\right)\,\right\}^{\,2}}
{2\,L_\mathrm{\,d}\,\left\{\,\left(\,\mathcal{P}_0\,\mathcal{P}_T\,A_\mathrm{\,out} \,-\,L_\mathrm{\, in}\,\right)\, \left(\,e^{\,\lambda_1\,\tau}\, - \, e^{\,\lambda_2\,\tau}\,\right)\, +\,L_\mathrm{\, d}\, \left(\,e^{\,\lambda_1\,\tau} \,+\, e^{\,\lambda_2\,\tau}\,\right)\,\right\}}} \,.\nonumber 
\end{eqnarray}

\subsection{An unpolarized target}
\label{subsec:An_unpolarized_target}

A special case of this system deserves comment.  Given that the beam is initially polarized what happens if the target is unpolarized?  One would imagine that the beam polarization should decrease and eventually reach zero.  We now analyze the equations of sect.~\ref{sec:Stored_beam_with_initial_polarization} when the target is unpolarized ({\it i.e.}\ $\mathcal{P}_T = 0$) and use the fact that $L_\mathrm{\,d} = \sqrt{\, \mathcal{P}_T^{\,2} \, A_\mathrm{\,out} \left( A_\mathrm{\,all}\, - \, K_\mathrm{\,in} \right) \, + \, L_\mathrm{\,in}^{\,2} } = L_\mathrm{\,in}$ when $\mathcal{P}_T = 0$ to obtain the beam polarization as a function of time
\begin{eqnarray}
\mathcal{P}(\tau)  & = & 
\mathcal{P}_0\ \displaystyle{e^{\,\left(\,\lambda_1 - \lambda_2\,\right)\ \tau}} \, ,
\end{eqnarray}
which is an exponentially decreasing function of $\tau$ for $\lambda_1 - \lambda_2 < 0$.  The beam polarization will not decrease in the special case of $\lambda_1 - \lambda_2 = 0$, but this only happens when $L_\mathrm{\,in} = 0$, {\it i.e.}\ when there is no depolarization.  The special case of $\lambda_1 - \lambda_2 = 0$, which does not lead to polarization buildup as seen from eq.~(\ref{eq:Initial_Polarization_J}), would be avoided by any experimental effort, thus is omitted from the rest of the discussion.  

The limit of beam polarization for large times when $\mathcal{P}_T = 0$ is
\begin{equation}
   \lim_{\tau \to \,\infty} \ \mathcal{P}(\tau) \ = 
\ 0\,.
\end{equation}
Thus, as expected, if the beam is initially polarized and the target unpolarized then the beam polarization will decrease with time and eventually the beam polarization will reduce to zero.  Thus a beam cannot gain polarization from an unpolarized target by spin filtering.

The figure of merit in this case simplifies to $\mathrm{FOM}(\tau) = N_0\,\mathcal{P}_0^{\,2}\,e^{\,\left(\,2\,\lambda_1 - \lambda_2\,\right)\,\tau}$ which is a monotonically decreasing function of $\tau$.  One can derive a polarization half-life in this case, the time taken for the polarization to decrease by a factor of $2$, by solving $\mathcal{P}(\tau) = \mathcal{P}_0\,/\,2$ to obtain
\begin{equation}
\tau_{\ \frac{1}{2}} \ = \ \frac{\ln 2}{\lambda_2 - \lambda_1} \ = \ \frac{\ln 2}{2\,n\,\nu\,L_\mathrm{\,d}} \,.
\end{equation}
This scenario occurs in an electron cooler, a device used to focus the beam in many storage rings.  The beam passes through a co-moving beam of unpolarized electrons with low transverse momentum, in order to dampen the transverse momentum of the antiprotons in the stored beam.  But the low electron areal densities in cooler beams, where typically $n \, \approx \, 10^{-18}~\mbox{fm}^{-2} \, = \, 10^{-19}~\mbox{mb}^{-1} $, causes the polarization half-life to be very large.  Thus this depolarization effect is negligible. 

It has recently been suggested that the positron-antiproton polarization transfer observable is very much enhanced at low energies \cite{Arenhovel:2007gi}.  This enhancement is the basis of the recent proposal by Walcher \emph{et al.}\ to polarize an antiproton beam by repeated interaction with a co-moving polarized positron beam in a storage ring \cite{Walcher:2007sj}.  All of the antiprotons remain within the beam in this scenario and one avoids the problem of the antiprotons annihilating with protons in an atomic gas target.  This large enhancement of the polarization transfer observable at low energies is due to the unlike charges of the positron and antiproton, and does not occur for the like charges case of antiproton-electron scattering.  Thus this does not effect the conclusion that depolarization of an antiproton beam in an electron cooler is negligible.

\subsection{A critical value for the target polarization}
\label{subsec:A_critical_value_for_the_target_polarization}

The beam polarization will also decrease for low values of the target polarization.  In fact there is a critical value of the target polarization $\mathcal{P}_T$ which keeps the beam polarization constant.  If the target polarization is above this critical value the polarization of the beam will increase, and if the target polarization is below this critical value the beam polarization will decrease.  The critical value is obtained by solving
\begin{equation}
\mathcal{P}(\tau) \ = \  \frac{L_\mathrm{\, d}\,\mathcal{P}_0 \, - \,\tanh\left(L_\mathrm{\, d}\,n\,\nu\,\tau\,\right)\,\left[\,L_\mathrm{\, in}\,\mathcal{P}_0 \, + \, \mathcal{P}_T\,\left(\,A_\mathrm{\,all} - K_\mathrm{\,in}\,\right)\,\right]}{L_\mathrm{\, d} \, + \, \tanh\left(L_\mathrm{\, d}\,n\,\nu\,\tau\,\right)\,\left[\,L_\mathrm{\, in} \, - \, \mathcal{P}_0\,\mathcal{P}_T\,A_\mathrm{\,out}\,\right]} \ = \ \mathcal{P}_0 \, ,
\end{equation}
for $\mathcal{P}_T$, where the time dependence will cancel leading to
\begin{equation}
\mathcal{P}_T^\mathrm{\,critical} \ = \ \frac{2\,\mathcal{P}_0\,L_\mathrm{\, in}}{\mathcal{P}_0^{\,2}\,A_\mathrm{\,out} - \left(\,A_\mathrm{\,all} - K_\mathrm{\,in}\,\right)}\,.
\end{equation}
For target polarizations below this critical value the maximum beam polarization occurs at time $\tau = 0$, and for target polarizations above this critical value the maximum beam polarization occurs at large times $\tau \rightarrow \infty$.

\section{Particles fed in for a limited time}
\label{sec:Particles_fed_in_for_a_limited_time}

The Heaviside step function could be used in the system of equations to explain the case of particles input into a beam for a certain amount of time after which the input is turned off and no more particles are fed into the beam, but spin filtering continues.  This scenario is under consideration by the PAX collaboration \cite{Barone:2005pu}.  The Heaviside function is a piecewise continuous function which is zero in one region and one everywhere else, it is used in many mathematical modeling problems to describe an external effect turned on or off after a certain duration of time.  In our case this external effect is the input of particles into the beam.  The Heaviside function is defined as  
\begin{eqnarray}
 H\left(\,\tau - \tau_c\,\right) \ = \ \left\{ 
\begin{array}{ll} 
0  & \hspace*{2em} \mbox{if} \ \tau < \tau_c \nonumber\\[2ex]  
1 & \hspace*{2em} \mbox{if} \ \tau \geq \tau_c

\end {array}
\right.\nonumber
\end{eqnarray}
and is used to describe an external effect turned on at time $\tau_c$, but we require an external effect on initially and turned off at time $\tau_c$, thus we need 
\begin{eqnarray}
H\left( \tau \right) - H \left(\,\tau - \tau_c\,\right) \ = \ \left\{ 
\begin{array}{ll} 
1  & \hspace*{2em} \mbox{if} \ 0 \leq \tau < \tau_c \nonumber \\[2ex] 0 & \hspace*{2em} \mbox{if} \ \tau \geq \tau_c
\end {array}
\right.\nonumber
\end{eqnarray}
Note that in the special case when $\tau_c = 0$, $\left[\,H\left( \tau \right) - H\left(\tau\right)\,\right]  = 0$, as this is in the second region.  This describes a physical situation where particles are being fed in for zero seconds, which is the same as saying no particles are fed in, so the second order ODE should be homogeneous in this case, which it is.  The extra term to add to the $\mathrm{d}\,N\,/\,\mathrm{d}\,\tau$ equation to account for particles being fed in at a constant rate $\beta$ per second for $\tau_c$ seconds after which the input is switched off is $\beta\,\left[\,H\left( \tau \right) - H \left(\,\tau - \tau_c\,\right)\,\right]$.

The equations are now broken into two pieces, {\it i.e.}\ discontinuous, but piecewise continuous.  The solutions will be broken into two regions $0\leq \tau < \tau_c$ and $\tau \geq \tau_c$\,, where the solutions in the region $0 \leq \tau < \tau_c$ should equal those in the inhomogeneous case presented in sect.~\ref{sec:Accumulation_of_antiprotons_in_the_ring}.  The initial conditions will be $N(0) = 0$ and $J(0) = 0$, thus $J'(0)= 0$.  The Heaviside function above is included in the second order ODE for $J(\tau)$ to obtain:
\begin{eqnarray}
& & \frac{\mathrm{d}^{\,2}\,J(\tau)}{\mathrm{d}\,\tau^2}  \ - \  \left(\,\lambda_1 + \lambda_2\,\right)\,\frac{\mathrm{d}\,J(\tau)}{\mathrm{d}\,\tau} \ + \ \lambda_1 \,\lambda_2\, J(\tau) \nonumber \\[1ex]
&  & \qquad \qquad \qquad = \ - \,n\, \nu \,\mathcal{P}_T\,\left(\,A_\mathrm{all} - K_\mathrm{in}\,\right) \, \beta\,\left[\,H\left( \tau \right) - H\left(\,\tau - \tau_c\,\right)\,\right] \, ,
\end{eqnarray}
which can be solved by Laplace Transform methods giving
\begin{eqnarray}
\label{eq:Heaviside_J_solution}
  \displaystyle{\frac{J(\tau)}{C_1}}  =  \left\{  \begin{array}{ll}
\lambda_1\, \left(\,1-e^{\,\lambda_2\,\tau}\,\right) \,+ \, \lambda_2\,\left(\,e^{\,\lambda_1\,\tau} - 1\,\right)
&  \ \ \ \mbox{if} \ 0 \leq \tau < \tau_c \\[3ex] 
\lambda_1\, \left(\,e^{\,\lambda_2\,\left(\,\tau - \tau_c\,\right)}\,-\, e^{\,\lambda_2\,\tau}\,\right) \,+ \, \lambda_2\,\left(\,e^{\,\lambda_1\,\tau} - e^{\,\lambda_1\,\left(\,\tau - \tau_c\,\right)}\,\right)
&  \ \ \ \mbox{if} \ \tau \geq \tau_c
\end {array}
\right.
\end{eqnarray}
where for convenience we have defined the constant factor
\begin{equation}
C_1 \ \equiv \ \frac{\beta\,\mathcal{P}_T\,\left(\,A_\mathrm{all} - K_\mathrm{in}\,\right)}{2\,L_\mathrm{d}\,\lambda_1\,\lambda_2} \,.
\end{equation}
One sees from eq.~(\ref{eq:Heaviside_J_solution}) that $J(\tau) = 0 \ \mbox{for all} \ \tau$ when $\tau_c = 0$, which is physically reasonable as there are never any particles in the beam if $\tau_c = 0$.  Also the $\mathcal{P}_T$ factor in $C_1$ indicates that $J(\tau)$ will always be zero if $\mathcal{P}_T = 0$ ({\it i.e.}\ if the target is unpolarized).  It can also be seen that the complete solution in the region $\tau \geq \tau_c$ is the combination of the solution in the region $0 \leq \tau < \tau_c$ and an additional part dependent on $\tau_c \,$; which is
\begin{equation}
C_1\,\left[\,\lambda_1\, \left(\,e^{\,\lambda_2\,\left(\,\tau - \tau_c\,\right)} - 1\,\right) \,+ \, \lambda_2\,\left(\,1-e^{\,\lambda_1\,\left(\,\tau - \tau_c\,\right)}\,\right)\,\right] \, , \nonumber
\end{equation}
and immediately one sees that when $\tau = \tau_c$ this additional part vanishes.  So when $\tau = \tau_c$ {\it i.e.}\ at the boundary between the two regions, the two solutions match.  Thus the solution for $J(\tau)$ is continuous as expected.  The expression for $J(\tau)$ in the first region $0 \leq \tau < \tau_c$  of eq.~(\ref{eq:Heaviside_J_solution}) is equal to the solution of the inhomogeneous system presented in eq.~(\ref{eq:Inhomogeneous_J_with_N0_0}).

A similar analysis as that done for $J(\tau)$ reveals the second order ODE for $N(\tau)$
\begin{eqnarray}
& & \frac{\mathrm{d}^{\,2}\,N(\tau)}{\mathrm{d}\,\tau^2}  \ - \  \left(\,\lambda_1 + \lambda_2\,\right)\,\frac{\mathrm{d}\,N(\tau)}{\mathrm{d}\,\tau} \ + \ \lambda_1 \,\lambda_2\, N(\tau) \nonumber \\[1ex]
&  & \qquad \qquad \qquad =   \,n\, \nu \,\left(\,I_\mathrm{all} - D_\mathrm{in}\,\right) \, \beta \,\left[\,H\left( \tau \right) - H\left(\,\tau - \tau_c\,\right)\,\right]\,.
\end{eqnarray}
The initial conditions are $N(0) = 0$ and $N'(0) = \beta\, \left[\,1 - H\left(\,-\tau_c\,\right)\,\right]$, the latter of which deserves comment.  The rate $N'(0)$ should be $\beta$ when $\tau_c > 0$ and $0$ when $\tau_c = 0$, corresponding to the case of particles being fed in for zero seconds, {\it i.e.}\ no particles fed in.  Note that $H\left(\,-\tau_c\,\right) = 1 \ \mbox{when} \ \tau_c = 0$ and $H\left(\,-\tau_c\,\right) = 0 \ \mbox{when} \ \tau_c > 0$, and note physically that $\tau_c$\,, the duration for which particles are fed into the beam, cannot be negative.  The Heaviside function in this initial condition forces the solution for $N(\tau)$ to be split into three regions.  On solving by Laplace Transform methods one obtains
\begin{eqnarray}
\label{eq:Heaviside_beam_intensity_solution}
\frac{N(\tau)}{C_2} =  \left\{  \begin{array}{ll} 
   \lambda_1 \,\left[\, 1- \left(\,\displaystyle{\frac{L_\mathrm{in} + L_\mathrm{d}}{I_\mathrm{all} - D_\mathrm{in}}}\,\right)\,e^{\,\lambda_2\,\tau} \,\right] \\[4ex]
\qquad \qquad + \ \lambda_2 \,\left[\,\left(\,\displaystyle{\frac{L_\mathrm{in} - L_\mathrm{d}}{I_\mathrm{all} - D_\mathrm{in}}}\,\right)\,e^{\,\lambda_1\,\tau}  - 1\,\right]
&  \mbox{if} \ 0 \leq \tau < \tau_c \\[9ex] 
\lambda_1 \,\left[\, e^{\,\lambda_2\,\left(\,\tau - \tau_c\,\right)}- \left(\,\displaystyle{\frac{L_\mathrm{in} + L_\mathrm{d}}{I_\mathrm{all} - D_\mathrm{in}}}\,\right)\,e^{\,\lambda_2\,\tau} \,\right] \\[4ex]
\qquad \qquad + \ \lambda_2 \,\left[\,\left(\,\displaystyle{\frac{L_\mathrm{in} - L_\mathrm{d}}{I_\mathrm{all} - D_\mathrm{in}}}\,\right)\,e^{\,\lambda_1\,\tau}  - e^{\,\lambda_1\,\left(\,\tau - \tau_c\,\right)}\,\right]
&  \mbox{if} \ \tau \geq \tau_c > 0 \\[7ex]
0 & \mbox{if} \ \tau_c = 0 
\end {array}
\right.
\end{eqnarray}
where again for convenience we have defined a constant factor
\begin{equation}
C_2 \ \equiv \ \frac{-\,\beta\,\left(\,I_\mathrm{all} - D_\mathrm{in}\,\right)}{2\,L_\mathrm{d}\,\lambda_1\,\lambda_2} \,.
\end{equation}
Again one sees that the complete solution in the region $\tau \geq \tau_c > 0$ is the combination of the solution in the region $0 \leq \tau < \tau_c$ plus an additional part dependent on $\tau_c \,$; which is
\begin{equation}
C_2\,\left[\,\lambda_1\, \left(\,e^{\,\lambda_2\,\left(\,\tau - \tau_c\,\right)} - 1\,\right) \,+ \, \lambda_2\,\left(\,1-e^{\,\lambda_1\,\left(\,\tau - \tau_c\,\right)}\,\right)\,\right] \,. \nonumber
\end{equation}
Immediately we see that when $\tau = \tau_c$ this additional part vanishes, thus the solution for $N(\tau)$ is continuous.  The solution in the first region $0 \leq \tau < \tau_c$ is equal to the solution from our inhomogeneous case presented in eq.~(\ref{eq:Inhomogeneous_N_with_N0_0}), and it satisfies the initial condition $N(0) = 0$.

We now present results for the polarization $\mathcal{P}(\tau) = J(\tau)\,/\,N(\tau)$ as a function of time, in both regions.  The polarization is undefined when there are no particles in the beam, thus we need not treat the case $\tau_c = 0$.  As expected in the $0 \leq \tau < \tau_c$ region $\mathcal{P}(\tau)$ equals the solution of our inhomogeneous case presented in eq.~(\ref{eq:Polarization_Inhomogeneous_N_equals_0}), and in the region $\tau \geq \tau_c > 0$ one finds
{\small
\begin{eqnarray}
\hspace*{-1.5em}  
& \hspace*{-1.5em}  & \mathcal{P}(\tau) \ =  \\[2ex]
& \hspace*{-1.5em} & \displaystyle{\frac{\mathcal{P}_T \,\left(\,A_\mathrm{all} - K_\mathrm{in}\,\right)\,\left[\,\lambda_1\,e^{\,\lambda_2\,\tau}\,\left(\,e^{-\lambda_2\,\tau_c}\,-\, 1\,\right) +  \lambda_2\,e^{\,\lambda_1\,\tau}\,\left(\,1 - e^{-\lambda_1\,\tau_c}\,\right)\,\right]
}{ \lambda_1 \,e^{\,\lambda_2\,\tau} \left[\, \left(\,I_\mathrm{all} - D_\mathrm{in}\,\right) e^{-\lambda_2\,\tau_c}- \left(\,L_\mathrm{in} + L_\mathrm{d}\,\right) \,\right] + \lambda_2 \,e^{\,\lambda_1\,\tau} \left[\,\left(\,L_\mathrm{in} - L_\mathrm{d}\,\right)  - \left(\,I_\mathrm{all} - D_\mathrm{in}\,\right) e^{-\lambda_1\,\tau_c}\,\right]} \nonumber \,.
}
\end{eqnarray}
}
The approximate initial rate of polarization buildup and the maximum polarization achievable will both reside in the $0 \leq \tau < \tau_c$ region, and thus will be identical to those presented in sect.~\ref{sec:Accumulation_of_antiprotons_in_the_ring}.  This is because the maximum polarization achievable occurs when the input rate is never switched off, {\it i.e.}\ in the $0 \leq \tau < \tau_c$ region.

\section{Summary}
\label{sec:Summary}

We have presented and solved systems of differential equations that describe the buildup of polarization of a beam by spin filtering in a storage ring, in various different scenarios.  These scenarios are: 1) spin filtering of a fully stored beam, 2) spin filtering while the beam is being accumulated {\it i.e.}\ unpolarized particles are continuously being fed into the beam, 3) the particle input rate is equal to the rate at which particles are being lost due to scattering beyond ring acceptance angle, thus the beam intensity remains constant, 4) increasing the initial polarization of a stored beam by spin filtering, 5) the input of particles into the beam is stopped after a certain amount of time, but the spin filtering continues.  The rate of depolarization of a stored polarized beam on passing through an electron cooler has also been shown to be negligible, because of the low electron areal densities in cooler beams.

The principal advantage of this method of polarization buildup is that it can be applied to antiprotons.  The investigation carried out in this paper will be of use to the PAX collaboration and to future projects that hope to produce a high intensity polarized antiproton beam, or to utilize spin filtering in a storage ring.  This work can also be applied to the recent idea of polarizing a beam by channeling through a bent crystal \cite{Ukhanov:2007tc}.

\pagebreak

{\small DOB would like to thank the Irish Research Council for Science, Engineering and Technology (IRCSET), for a postgraduate research scholarship.  NHB is grateful to Enterprise Ireland for the award of a grant under the International Collaboration Program to facilitate a visit to INFN at the University of Torino.  We are also grateful to M.~Anselmino, E.~Leader, N.~N.~Nikolaev, F.~Rathmann and W.~Vogelsang for helpful comments.}

\end{document}